\documentclass[lettersize,journal]{IEEEtran}
\IEEEoverridecommandlockouts
\usepackage{cite}
\usepackage{algorithm}
\usepackage{algorithmic}
\usepackage{amsmath,amssymb,amsfonts}
\usepackage{graphicx}
\usepackage{textcomp}
\usepackage{xcolor}
\usepackage[center]{subfigure}
\usepackage{multirow}
\usepackage{bm}

\newcommand{\reffig}[1]{Fig.~\ref{#1}}
\newcommand{\reftab}[1]{Table~\ref{#1}}
\newcommand{\refsec}[1]{Section~\ref{#1}}
\newcommand{\refeq}[1]{(\ref{#1})}
\newcommand{\tabincell}[2]{\begin{tabular}{@{}#1@{}}#2\end{tabular}}

\begin{document}

\title{An Architectural Error Metric for CNN-Oriented Approximate Multipliers}

\author{Ao Liu, Jie Han,~\IEEEmembership{Senior Member,~IEEE,} Qin Wang, Zhigang Mao, and Honglan Jiang,~\IEEEmembership{Member,~IEEE}
\thanks{A. Liu, Q. Wang, Z. Mao, H. Jiang are with the Department of Micro-Nano Electronics, Shanghai Jiao Tong University, Shanghai 200240, China. 
(e-mail: 2017la@sjtu.edu.cn, qinqinwang@sjtu.edu.cn, maozhigang@sjtu.edu.cn, honglan@sjtu.edu.cn)
}
\thanks{J. Han is with the Department of Electrical and Computer Engineering, University of Alberta, Edmonton, AB T6G 1H9, Canada. 
(e-mail: jhan8@ualberta.ca)
}}

\maketitle

\begin{abstract}
As a potential alternative for implementing the large number of multiplications in convolutional neural networks (CNNs), approximate multipliers (AMs) promise both high hardware efficiency and accuracy. However, the characterization of accuracy and design of appropriate AMs are critical to an AM-based CNN (AM-CNN). In this work, the generation and propagation of errors in an AM-CNN are analyzed by considering the CNN architecture. Based on this analysis, a novel AM error metric is proposed to evaluate the accuracy degradation of an AM-CNN, denoted as the architectural mean error (AME). The effectiveness of the AME is assessed in VGG and ResNet on CIFAR-10, CIFAR-100, and ImageNet datasets. Experimental results show that AME exhibits a strong correlation with the accuracy of AM-CNNs, outperforming the other AM error metrics. To predict the accuracy of AM-CNNs, quadratic regression models are constructed based on the AME; the predictions show an average of 3\% deviation from the ground-truth values. Compared with a GPU-based simulation, the AME-based prediction is about $10^{6}\times$ faster.
\end{abstract}


\section{Introduction}
With the development of artificial intelligence, the number of parameters and arithmetic operations of neural networks increases rapidly. How to deploy large-scale neural network applications to hardware platforms with limited resources and power consumption has become a crucial issue. Leveraging the error tolerance inherent in convolutional neural networks (CNNs), approximate computing has emerged as a promising technique to enhance the efficiency of CNN accelerators~\cite{armeniakos2022hardware, guella2024marlin}. As multiplication dominates the computations in CNNs, a direct and effective approach for approximate computing in CNNs is utilizing approximate multipliers (AMs)~\cite{jiang2020approximate, mrazek2019alwann, tasoulas2020weight, zervakis2021control, spantidi2021positive}. In this case, the key question is how to select or design appropriate AMs for a CNN under a certain accuracy constraint.

In general, the accuracy of an AM-based CNN (AM-CNN) is assessed through simulations using frameworks such as TFApprox~\cite{vaverka2020tfapprox}, ProxSim~\cite{de2020proxsim}, AdaPT~\cite{danopoulos2022adapt}, and ApproxTrain~\cite{gong2023approxtrain}. In these frameworks, approximate multiplication is implemented by accessing the look up table (LUT) containing the products of all possible input combinations. Based on this, the 2D convolution operation in the TensorFlow or PyTorch framework is modified to support simulations for AM-CNNs. However, simulations provide only accuracy, lacking the capacity to guide the AM selection or design. Moreover, even using state-of-the-art simulation approaches, the search for the optimal AM in a large design space is still time-consuming.

An alternative approach to evaluating the accuracy of an AM-CNN involves accuracy prediction. This method disregards the specific hardware structure of the AM and focuses solely on its error characteristics. For example, treating the errors introduced by AMs as noise~\cite{hanif2018error, hammad2018impact, de2021efficient}, or representing an AM by its error metrics~\cite{ansari2019improving, mo2023learning, kim2021effects, spantidi2022much}. By modeling the impact of the AM error characteristics on the AM-CNN accuracy, guidance can be provided for CNN-oriented AM selection and design. However, although offering theoretical insights, many existing accuracy prediction methods for AM-CNN have limited practical applicability. The main challenges are as follows.
\begin{itemize}
    \item CNNs generally contain numerous multiplications and nonlinear functions, making the error propagation process within AM-CNNs difficult to mathematically model. Consequently, current accuracy prediction methods for AM-CNNs are often rather complex.
    \item A universally accepted and efficient method to characterize the error characteristics of a CNN-oriented AM is lacking. Specifically, there is no error metric for AMs that can adequately captures their impacts on the accuracy of AM-CNNs.
\end{itemize}

In this work, we linearly model the process of error generation and propagation within AM-CNNs. Based on this model, a novel AM error metric denoted as the architectural mean error (AME) is proposed to evaluate the accuracy degradation of an AM-CNN. The experimental results show that the AME exhibits a strong correlation with the accuracy of AM-CNNs, and the AME-based quadratic regression models can predict the accuracy of AM-CNNs with an average of 3\% deviation. Given an AM-CNN, computing the AME involves only a few matrix operations, making the AME-based accuracy prediction about $10^{6}\times$ faster than a GPU-based simulation.

The novel contributions of this work are as follows.
\begin{itemize}
    \item A linear model is developed to fit the error generation and propagation process within an AM-CNN.
    \item An architectural error metric is proposed for AM to evaluate the accuracy degradation of AM-CNNs.
    \item Quadratic regression models are trained based on the AME to predict the accuracy of AM-CNNs.
    \item The effectiveness of the proposed AME is verified by using it to accelerate the selection process of the Pareto-optimal AMs for CNN applications.
\end{itemize}

The rest of the paper is organized as follows. \refsec{sec:related} introduces the studies related to this work. \refsec{sec:error_metric} shows the definitions of the error metrics and outlines our motivation. \refsec{sec:AME} describes the process for deriving the proposed AME. \refsec{sec:exp} presents the experimental results for evaluating the effectiveness of AME. \refsec{sec:application} provides application cases of AME. \refsec{sec:discussion} discusses the characteristics of AME. Finally, \refsec{sec:conclusion} concludes the paper.

\section{Related Work}
\label{sec:related}
To explore the relationship between the error metrics of an AM and the accuracy of the corresponding AM-CNN, various approaches have been explored in state-of-the-art researches.

Ansari et al.~\cite{ansari2019improving} trained two-class classifiers to categorize AMs based on whether they degrade the accuracy of AM-CNNs. These classifiers take various AM error metrics as inputs; these metrics are subsequently ranked based on their significance in achieving a high classification accuracy. The experimental results highlighted two critical AM error metrics, the variance of error (VarE) and the root-mean-square error (RMSE). However, this study only analyzed a small-scale CNN (i.e., the LeNet-5) on small datasets such as MNIST and SVHN. In addition, the classification accuracy of the constructed AM classifiers is pretty low, e.g., the highest accuracy is below 86\%.

Similarly, \cite{mo2023learning} trained not only classifiers but also regressors to predict the AM-CNN accuracy based on AM error metrics. More complex CNNs such as VGG and ResNet, are involved in this study. By performing a dropout feature ranking, \cite{mo2023learning} finally summarized three important AM error metrics, the mean error (ME), the mean error distance (MED), and the error rate (ER). Although high accuracy is achieved, the classifiers and regressors constructed in \cite{mo2023learning} are generally complex and require a large number of samples for the training process. Moreover, this study lacks a mathematical explanation for the effectiveness of these metrics.

Kim et al.~\cite{kim2021effects} analyzed the error generation and propagation process in an AM-CNN, demonstrating that AMs with low values of variance of relative error (VarRE) are unlikely to lead to accuracy degradation in AM-CNNs. However, in the experiments, $32\times 32$ fixed-point multiplications are utilized in the inference of the considered CNN models, which is far more accurate than the commonly used $8\times 8$ designs. Thus, the error effects of the AMs on the AM-CNNs cannot be effectively illustrated. Also, only three AMs are tested to verify the analyses, the applicability of the conclusions to arbitrary AMs has not been demonstrated.

\section{Analysis of Error Metrics}
\label{sec:error_metric}
\subsection{Error Metric Definitions}
Errors introduced by AMs are propagated and accumulated in an AM-CNN, causing feature changes and thus affecting the accuracy. Both the output errors of an AM and the accumulated errors in the AM-CNN can be characterized by error metrics. To distinguish these two types of errors, we denote the former as $\varepsilon$ and the latter as $e$; they are given by
\begin{equation}
    \begin{aligned}
        \varepsilon &= a'-a\;,\\
        e &= y'-y\;,
    \end{aligned}
    \label{eq:error}
\end{equation}
where $a'$ and $a$ are the approximate and the corresponding accurate multiplication results, respectively; $y'$ and $y$ are the approximate and the corresponding accurate output features of a neuron, respectively. Also, to evaluate the relative deviation from the accurate value, the relative errors ($\varepsilon_r$ and $e_r$) are defined. They are given by
\begin{equation}
    \begin{aligned}
        \varepsilon_r &= \frac{a'-a}{a}\;,\\
        e_r &= \frac{y'-y}{y}\;.
    \end{aligned}
    \label{eq:relative_error}
\end{equation}
By calculating the probability ($p$), mean ($\mu$), variance ($\sigma^2$), and other measurements of errors and relative errors, various error metrics are obtained. \reftab{tab:EMetrics} shows the definitions of some common AM error metrics, where $n$ and $n_r$ are the total numbers of errors and relative errors, respectively. The error metrics for AM-CNNs are defined similarly, with $\varepsilon$ and $\varepsilon_r$ in \reftab{tab:EMetrics} replaced by $e$ and $e_r$, respectively, while retaining the same metric names.

\begin{table}[t]
    \caption{The definitions of AM error metrics.}
    \begin{center}
        \scalebox{1.0}{
        \begin{tabular}{lll}
            \hline
            \textbf{Metric} & \textbf{Description} & \textbf{Formula} \\
            \hline
            \textbf{ER} & Error rate. & $p(\varepsilon \ne 0)$ \\
            \textbf{ME} & Mean error. & $\mu(\varepsilon)$ \\ 
            \textbf{MRE} & Mean relative error. & $\mu(\varepsilon_r)$ \\
            \textbf{MED} & Mean error distance. & $\mu(|\varepsilon|)$ \\
            \textbf{MRED} & Mean relative error distance. & $\mu(|\varepsilon_r|)$ \\
            \textbf{VarE} & Variance of error. & $\sigma^2(\varepsilon)$ \\
            \textbf{VarRE} & Variance of relative error. & $\sigma^2(\varepsilon_r)$ \\
            \textbf{VarED} & Variance of error distance. & $\sigma^2(|\varepsilon|)$ \\
            \textbf{VarRED} & Variance of relative error distance. & $\sigma^2(|\varepsilon_r|)$\\
            \textbf{MSE} & Mean-square error. & $\mu({\varepsilon}^2)$ \\
            \textbf{RMSE} & Root-mean-square error. & $\sqrt{\mu({\varepsilon}^2)}$ \\ 
            \textbf{WCE} & Worst-case error. & $\max_{i\in[1,n]}{|\varepsilon_i|}$\\
            \textbf{WCRE} & Worst-case relative error. & $\max_{i\in[1,n_r]}{|\varepsilon_{r_i}|}$\\
            \hline
        \end{tabular}}
        \label{tab:EMetrics}
    \end{center}
\end{table}

To enhance hardware efficiency, the data in CNN is commonly quantized to a lower bit width, such as 8-bit~\cite{armeniakos2022hardware}. For an AM with inputs of low bit width, its error metrics can be evaluated through traversing all possible input combinations. For example, the mean error (ME) of an $8 \times 8$ unsigned AM can be calculated as follows
\begin{equation}
    \mu(\varepsilon) = \sum_{x=0}^{2^{8}-1}\sum_{w=0}^{2^{8}-1}d(x, w) {\varepsilon}(x, w)\;, 
    \label{eq:ME_dist}
\end{equation}
where $d(x, w)$ is the joint probability mass function of the inputs, and ${\varepsilon}(x, w)$ is the $\varepsilon$ for $x \times w$. To simplify the expression, \refeq{eq:ME_dist} can be transformed to
\begin{equation}
    \mu(\varepsilon) = \langle \mathbf{D},\, \mathbf{\Delta}\rangle_{\rm F}\;, 
    \label{eq:ME_matrix}
\end{equation}
where $\langle \, \,  \rangle_{\rm F}$ is the Frobenius inner product operation (i.e., the element-wise inner product of two matrices); $\mathbf{D}$ and $\mathbf{\Delta}$ are two $2^{8} \times 2^{8}$ matrices containing the $d(x, w)$ and $\varepsilon(x,w)$ for all possible $x \times w$, respectively. As $\mathbf{\Delta}$ contains all possible output errors of an AM, it can be referred to as the error matrix of the AM.

\subsection{Motivation: Error Metrics and Accuracy of AM-CNNs}
Given an AM-based application, it is feasible to estimate the output error metrics of the application based on the error metrics of the employed AMs. This approach requires analyzing and modeling the error generation and propagation process within the application~\cite{castro2018compiler, vaeztourshizi2023efficient}. As the quality of an AM-CNN is generally evaluated by the degradation of accuracy (e.g., classification accuracy) rather than the output error, our focus shifts to the error metrics associated with the output features of the last approximate NN layer.

Since each CNN layer outputs multiple features, individually analyzing the impact of changes in each feature on accuracy is less meaningful. Hence, we employ the error statistics across all output features of a layer as the error metrics of the layer, e.g., the ME of a layer is given by
\begin{equation}
   \mu(e) = \frac{1}{SMHW}\sum_{i=1}^{S}\sum_{j=1}^{M}\sum_{k=1}^{H}\sum_{l=1}^{W}e_{i,j,k,l}\;,
    \label{eq:ME_layer}
\end{equation}
where $S$ is the total number of test cases; $H$, $W$, and $M$ denote the height, width and channel number of the output feature maps of this layer, respectively.

We conducted experiments to assess the correlation between the accuracy and the error metrics of the last approximate layer for AM-CNNs. The experiments cover thousands of AM-CNNs given in \refsec{sec:exp}. \reftab{tab:correlation} reports the Pearson correlation coefficients (PCCs), calculated as per \refeq{eq:pcc}, for various configurations of AM-based VGG-16 and ResNet-18 on the CIFAR-10 dataset.
\begin{equation}
     \rho = \frac{{\rm Cov}(Acc, Metric)}{\sigma(Acc)\sigma(Metric)}\;,
    \label{eq:pcc}
\end{equation}
where $Acc$ is the set of classification accuracy of AM-CNNs, and $Metric$ represents the set of error metrics corresponding to these AM-CNNs. Note that, unlike other non-negative metrics, the values of ME and mean relative error (MRE) can be either positive or negative. This characteristic leads to conflicting correlations between these values and the AM-CNN accuracy. For example, both higher negative ME and lower positive ME (i.e., MEs closer to 0) can indicate smaller errors and are generally associated with higher AM-CNN accuracy. Hence, the PCCs for positive and negative values of ME and MRE are calculated separately, as shown in \reftab{tab:correlation}.

\begin{table}[h]
    \caption{PCCs between the classification accuracy and the error metrics of the last approximate layer for AM-CNNs.}
    \begin{center}
        \scalebox{1.0}{
        \begin{tabular}{lcc}
            \hline
            \textbf{Metric} & \textbf{VGG-16} & \textbf{ResNet-18} \\
            \hline
            \textbf{ER} & 0.046 & 0.511\\
            \textbf{ME} & \textbf{-0.844} $|$ \textbf{0.696} $^{\mathrm{a}}$ & \textbf{-0.955} $|$ \textbf{0.854} $^{\mathrm{a}}$ \\ 
            \textbf{MRE} & -0.743 $|$ 0.310 $^{\mathrm{a}}$ & -0.906 $|$ 0.780 $^{\mathrm{a}}$ \\
            \textbf{MED} & -0.566 & -0.761 \\
            \textbf{MRED} & -0.505 & -0.597 \\
            \textbf{VarE} & -0.488 & -0.771 \\
            \textbf{VarRE} & -0.465 & -0.426 \\
            \textbf{VarED} & -0.487 & -0.741 \\
            \textbf{VarRED} & -0.465 & -0.426\\
            \textbf{MSE} & -0.483 & -0.705 \\
            \textbf{RMSE} & -0.583 & -0.780 \\ 
            \textbf{WCE} & -0.615 & -0.852\\
            \textbf{WCRE} & -0.509 & -0.513\\
            \hline
            \multicolumn{3}{l}{$^{\mathrm{a}}$PCC for positive values $|$ PCC for negative values.}
        \end{tabular}}
    \label{tab:correlation}
    \end{center}
\end{table}

It can be seen that the ME, essentially revealing the error expectation for the last approximate layer, exhibits the strongest correlation with the AM-CNN accuracy.
Hence, we propose to construct a new AM error metric to estimate the error expectation of the last approximate layer. This metric, presented in the form of \refeq{eq:ME_matrix}, should have a strong correlation with the AM-CNN accuracy.

\section{Proposed Architectural Error Metric}
\label{sec:AME}
This section first constructs a linear model based on two assumptions to fit the error generation and propagation process in an AM-CNN. A novel AM error metric is then proposed, which contains the information pertaining to both the CNN architecture and the AM error characteristics. 

\subsection{Error Generation}
As one of the fundamental and computationally intensive components in CNNs, the convolutional layer typically accounts for the majority of computations. Therefore, designing AM-CNN generally involves approximating the convolutional layer~\cite{mrazek2019alwann, tasoulas2020weight, zervakis2021control, spantidi2021positive}. The calculation for an output feature ($y$) of the convolutional layer is expressed as
\begin{equation}
    y=\sum_{i=1}^{N}x_iw_i+b\;,
    \label{eq:layer}
\end{equation}
where $N$ is the total number of weights in a filter, $x_i$ is the $i$th input activation, $w_i$ is the weight for $x_i$, and $b$ is the bias. According to \refeq{eq:error}, the corresponding approximate output feature ($y'$) of the layer with multiplications implemented by the same type of AMs can be described as
\begin{equation}
    y' = \sum_{i=1}^{N}(x_iw_i+\varepsilon(x_i,w_i))+b\;,
    \label{eq:approx_layer}
\end{equation}
and the error of this output feature is given by
\begin{equation}
    e_{out} = y'-y = \sum_{i=1}^{N}\varepsilon(x_i,w_i)\;.
    \label{eq:output_error}
\end{equation}
The expectation of $e_{out}$ can be obtained as
\begin{equation}
   {\rm E}(e_{out}) = {\rm E}(\sum_{i=1}^{N}\varepsilon(x_i,w_i)) = \sum_{i=1}^{N}{\rm E}(\varepsilon(x_i,w_i))\;.
    \label{eq:E_output_error}
\end{equation}
Given \refeq{eq:ME_layer}, the error expectation of this layer is calculated as
\begin{equation}
   {\rm E}(e) = \frac{1}{MHW}\sum_{j=1}^{M}\sum_{k=1}^{H}\sum_{l=1}^{W}{\rm E}(e_{out_{j,k,l}})\;.
    \label{eq:E_mean_error}
\end{equation}

\textbf{Assumption 1: ${\rm E}(e)$ can be estimated as per the data distribution}. During the entire convolution process, each weight multiplies with almost every input activation in the same channel. Moreover, there is a certain correlation between different channels. Hence, ${\rm E}(e)$ can be estimated as 
\begin{equation}
    {\rm E}(e) \approx \langle N {\boldsymbol p}^\mathrm{T} {\boldsymbol f} ,\, \mathbf{\Delta} \rangle_{\rm F}\;,
   \label{eq:E_estimate}
\end{equation}
where ${\boldsymbol p}$ and ${\boldsymbol f}$ are two row vectors, ${\boldsymbol p}$ contains the probabilities ($p$) for all possible values of $x$, while ${\boldsymbol f}$ contains the frequencies ($f$) for all possible values of $w$; and $\mathbf{\Delta}$ is the AM error matrix as in \refeq{eq:ME_matrix}. For the case of using $8 \times 8$ unsigned AM, there are ${\boldsymbol p}=(p(x=0),\,p(x=1),\,\cdots,\,p(x=2^{8}-1))$, and ${\boldsymbol f}=(f(w=0),\,f(w=1),\,\cdots,\,f(w=2^{8}-1))$. Note that for an AM-CNN without weight fine-tuning (e.g., retraining), ${\boldsymbol f}$ is fixed and the same as that of the original exact CNN model, while ${\boldsymbol p}$ is determined by the input activation distribution of this layer.

\begin{figure}[t]
    \begin{center}
        \includegraphics [width=0.9\linewidth]{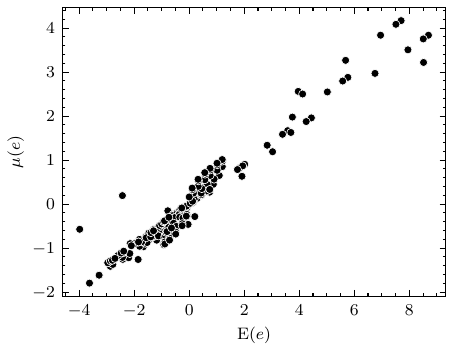}
        \caption{Scatterplot of AM-based convolutional layers, where the abscissa is the ${\rm E}(e)$ estimated by using \refeq{eq:E_estimate}, and the ordinate is the $\mu(e)$ measured by using \refeq{eq:ME_layer}.}
        \label{fig:single_layer}
    \end{center}
\end{figure}

Approaches similar to this assumption (i.e., predicting the error based on the data distribution) have been utilized in previous works, yielding good results~\cite{tasoulas2020weight, zervakis2021control}. To further verify the feasibility of this assumption, experiments were conducted to compare the ${\rm E}(e)$ estimated by using \refeq{eq:E_estimate} with the $\mu(e)$ measured by using \refeq{eq:ME_layer}. These experiments were conducted on the last convolutional layer of VGG-16 using the CIFAR-10 dataset, involving hundreds of AMs given in \refsec{sec:exp}. The experimental results are presented in \reffig{fig:single_layer}. It demonstrates that, despite a numerical deviation, the estimated ${\rm E}(e)$ exhibits an extremely strong correlation with the measured $\mu(e)$ (with a PCC exceeding 0.99). This result is sufficient for error estimation as we prioritizes the difference between the errors caused by different AMs, over the exact error values.

\subsection{Error Propagation}
In an AM-CNN, only the first approximate layer and the layers before it have accurate inputs, while all the other layers are influenced by the output errors of their previous approximate layers. An output feature of the AM-based convolutional layer with inaccurate inputs is given by
\begin{equation}
    y''=\sum_{i=1}^{N}((x_i+e_{in_i})w_i+\varepsilon(x_i+e_{in_i},w_i))+b\;,
    \label{eq:sub_approx_layer}
\end{equation}
where $e_{in_i}$ is the input error for the $i$th input activation. The error of this output feature is calculated as
\begin{equation}
    e'_{out} = y''-y = \sum_{i=1}^{N} e_{in_i}w_i + \varepsilon(x_i+e_{in_i},w_i)\;.
    \label{eq:sub_output_error}
\end{equation}
The expectation of $e'_{out}$ is given by
\begin{equation}
    {\rm E}(e'_{out}) = {\rm E}(\sum_{i=1}^{N} e_{in_i}w_i) + {\rm E}(\sum_{i=1}^{N}\varepsilon(x_i+e_{in_i},w_i))\;.
    \label{eq:E_sub_output_error}
\end{equation}
It can be seen that the error of this output feature is derived from two mutually independent sources, i.e., the input errors ($e_{in}$) and the AM output errors ($\varepsilon$).

\textbf{Assumption 2: ${\rm E}(e'_{out})$ can be fitted as a linear combination of the error expectations of the input and AM}. Denote this layer as layer $t$, and the ${\rm E}(e)$ (defined in \refeq{eq:E_mean_error}) corresponding to this layer and the previous layer as ${\rm E}(e_t)$ and ${\rm E}(e_{t-1})$, respectively. The ${\rm E}(e'_{out})$ is fitted as
\begin{equation}
    {\rm E}(e'_{out}) \approx \alpha_t{\rm E}(e_{t-1}) + {\rm E}(\sum_{i=1}^{N}\varepsilon(x_i,w_i))\;,
    \label{eq:estimate_E_sub}
\end{equation}
where $\alpha_t$ is the error propagation factor for layer $t$. The last term of \refeq{eq:estimate_E_sub} is equal to \refeq{eq:E_output_error}, which is the error expectation under accurate inputs.

Then, the ${\rm E}(e_t)$ is given by
\begin{equation}
    {\rm E}(e_{t}) \approx \alpha_t{\rm E}(e_{t-1}) + {\rm E}(e_{t,0})\;,
    \label{eq:linear_model}
\end{equation}
where ${\rm E}(e_{t,0})$ denotes the intrinsic error expectation of layer $t$, which can be estimated by using \refeq{eq:E_estimate}.

\begin{figure}[t]
    \begin{center}
        \includegraphics [width=1.0\linewidth]{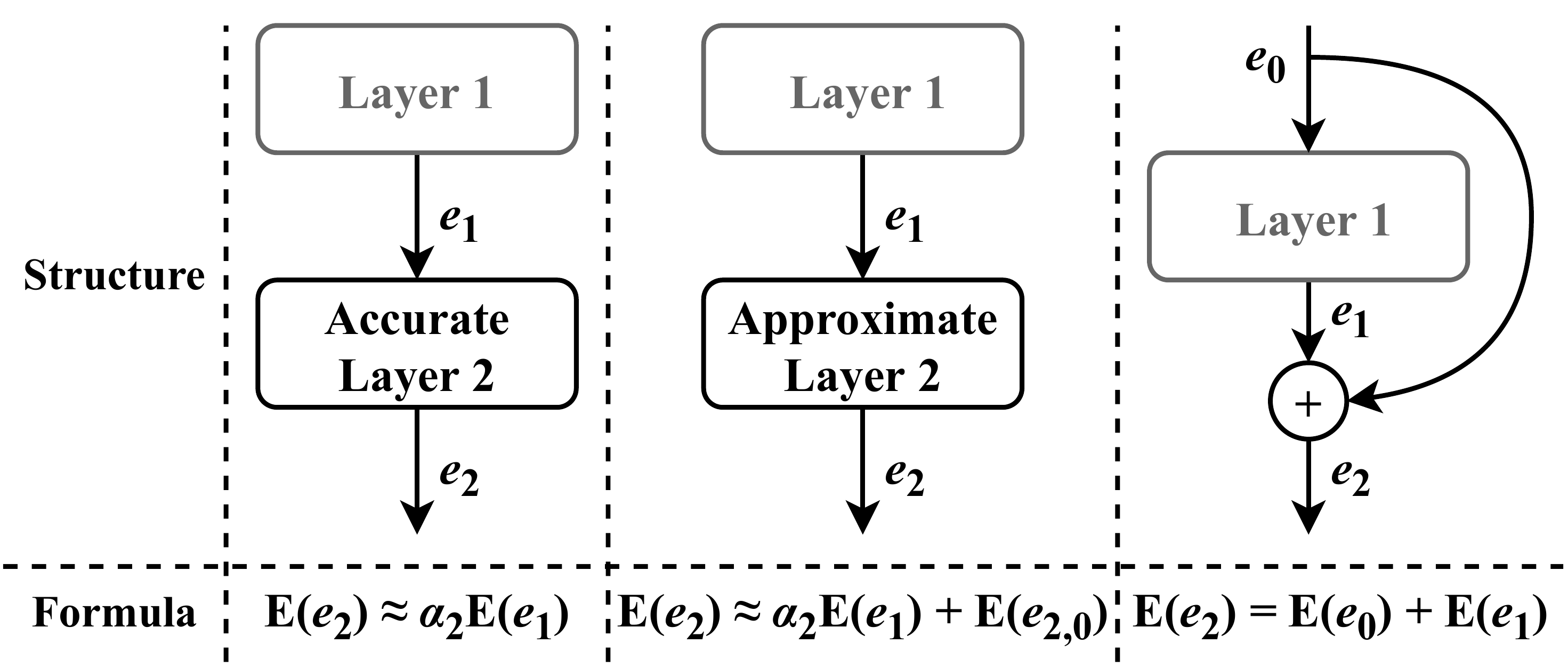}
        \caption{The proposed linear error propagation model.}
        \label{fig:error_prop}
    \end{center}
\end{figure}

By extending \refeq{eq:linear_model}, the linear error propagation model can be obtained, as shown in \reffig{fig:error_prop}. The accurate layer in \reffig{fig:error_prop} can be a convolutional layer or a batch normalization layer, while the approximate layer is an AM-based convolutional layer. Although this model is somewhat oversimplified, it is efficient enough. \textbf{Our goal is to develop a more efficient method for evaluating the accuracy of AM-CNNs than traditional simulations.} Under this premise, any complex error propagation model that makes the error estimation process slower than simulation is meaningless. Details of employing this model to fit the error propagation within various CNN layers are as follows.

\subsubsection{Convolutional Layer}
\label{subsubsec:conv}
Based on \refeq{eq:linear_model}, the error propagation process within an exact or approximate convolutional layers can be linearly fitted by choosing an appropriate error propagation factor $\alpha$. The $\alpha$ quantifies the impact of the input errors on the output errors for a layer; however, it is hard to obtain the corresponding mathematical formula. According to \refeq{eq:E_sub_output_error}, when $e_{in_i}\ll x_i$, the error primarily propagates from an input to the output through multiplication by the fixed weight ($w_i$). Therefore, it is feasible to treat $\alpha$ as a constant for a pretrained CNN layer and approximate it by empirical values obtained in experiments. Specifically, for a layer, after measuring the mean values of the input, output, and intrinsic errors separately on a given dataset, $\alpha$ can be calculated based on the formulas in \reffig{fig:error_prop}. Moreover, several principles can be applied to obtain a better $\alpha$, as follows.
\begin{itemize}
    \item $\alpha$ is expected to be independent of the AM, and depends solely on the CNN. However, since this is an idealized approximate model, the actual measured $\alpha$ may vary slightly depending on the AM used.
    \item A suitable AM for measuring $\alpha$ should degrade the accuracy of the corresponding AM-CNN by 2\% to 10\%.
    \item The absolute value of valid $\alpha$ is usually less than 10, and more often close to 1.
    \item To take into account AMs with various error characteristics, it is better to measure the $\alpha$ for multiple typical AMs and then take the average.
\end{itemize}

\begin{figure}[t]
    \begin{center}
        \includegraphics [width=1.0\linewidth]{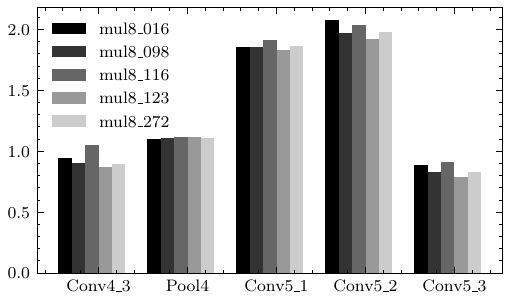}
        \caption{$\alpha$ values for four AM-based approximate convolutional layers and a max pooling layer in VGG-16 on CIFAR-10.}
        \label{fig:alpha}
    \end{center}
\end{figure}

\reffig{fig:alpha} shows the measured $\alpha$ of some layers for several AM-based VGG-16 on CIFAR-10 dataset, where the AMs were selected from EvoApprox8b library~\cite{mrazek2017evoapprox8b}.

\subsubsection{Nonlinear Layer}
Nonlinear operations in CNNs typically involve the ReLU activation function and pooling. A shared characteristic of these operations is the ability to selectively extract specific significant features while leaving their values unchanged. Specifically, ReLU extracts features with positive values, while max pooling extracts features with relatively high values. Therefore, when focusing exclusively on the error expectation corresponding to the significant features, nonlinear layers can be ignored. For example, if we only count the errors on the positive features when measuring the error propagation factor $\alpha$, the ReLU activation function can be ignored.

A verification of the above discussion is shown in \reffig{fig:alpha}. We measured the $\alpha$ of a max pooling layer (denoted as Pool4), which is the ratio of the mean values of the input and output errors. This measurement focus on the errors on positive features, and the resultant $\alpha$ is close to 1. Hence, the error expectations for the input and output features of the layer are identical, allowing to discard the impacts of this layer on error propagation.

\subsubsection{Batch Normalization Layer}
The batch normalization layer normalizes its inputs as
\begin{equation}
    y = \gamma \frac{x-\mu(x)}{\sqrt{\sigma^2(x)+\delta}} + \beta \;,
    \label{eq:BN}
\end{equation}
where $\mu(x)$ and $\sigma^2(x)$ are the mean and variance of the batch of inputs, respectively; $\delta$ is a small constant utilized to prevent division by zero; $\gamma$ and $\beta$ are the scaling and offset factors, respectively. It is noteworthy that these parameters ($\delta$, $\mu(x)$, $\sigma^2(x)$, $\gamma$, $\beta$) are determined during CNN construction and training, and remain fixed during inference. Therefore, for a batch normalization layer with inaccurate inputs, the calculation during the inference process is as follows.
\begin{equation}
    y' = \gamma \frac{x+e_{in}-\mu(x)}{\sqrt{\sigma^2(x)+\delta}} + \beta \;.
    \label{eq:BN_app}
\end{equation}
Thus, the output error of the batch normalization layer is given by
\begin{equation}
    e_{out} = y'-y = \frac{\gamma}{\sqrt{\sigma^2(x)+\delta}} e_{in} \;.
    \label{eq:BN_prop}
\end{equation}
\refeq{eq:BN_prop} illustrates that during the inference process, the input error $e_{in}$ of a batch normalization layer is propagated linearly to the output. This is consistent with the proposed error propagation model for exact convolutional layer, and therefore can also be represented by the leftmost column of \reffig{fig:error_prop}.

\subsubsection{Residual Layer}
According to the linearity of the expectation, the error expectations are added in the residual layer implemented by addition, as shown in the rightmost column of \reffig{fig:error_prop}.

\subsection{The Architectural Mean Error}
Based on the proposed linear model, the error expectation of the last approximate layer can be recursively estimated. An example is presented below to explain this in detail.

\begin{figure}[t]
    \begin{center}
        \includegraphics [width=1.0\linewidth]{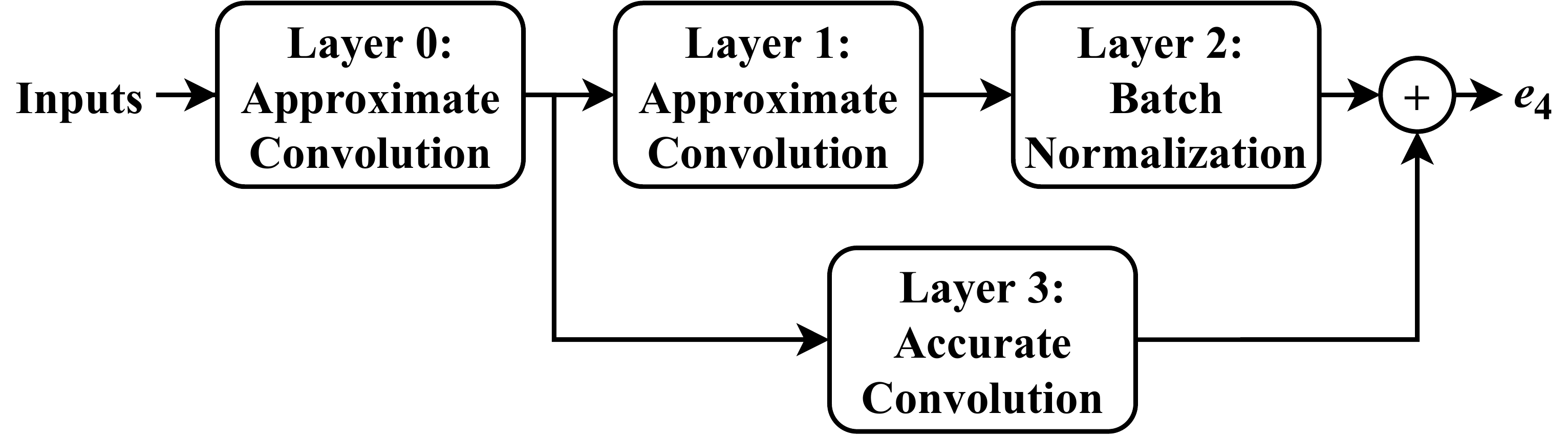}       
        \caption{Example of error estimation.}
        \label{fig:prop_example}
    \end{center}
\end{figure}

\reffig{fig:prop_example} shows a typical part of an AM-CNN. Based on the proposed error propagation model (\reffig{fig:error_prop}), when the inputs to this part are accurate, the ${\rm E}(e_4)$ can be fitted as
\begin{equation}
    \begin{aligned}
        {\rm E}(e_4) &\approx \alpha_2 (\alpha_1 {\rm E}(e_{0,0}) + {\rm E}(e_{1,0})) + \alpha_3 {\rm E}(e_{0,0}) \\ &= (\alpha_1 \alpha_2 + \alpha_3) {\rm E}(e_{0,0}) + \alpha_2 {\rm E}(e_{1,0}) \;.
    \end{aligned}
    \label{eq:prop_example}
\end{equation}
By using \refeq{eq:E_estimate}, ${\rm E}(e_4)$ can be further estimated as
\begin{equation}
    \begin{aligned}
        {\rm E}(e_4) \approx& \langle (\alpha_1 \alpha_2 + \alpha_3) N_0 {\boldsymbol p_0}^\mathrm{T} {\boldsymbol f_0},\, \mathbf{\Delta}_0 \rangle_{\rm F} \\ &+ \langle \alpha_2 N_1{\boldsymbol p_1}^\mathrm{T} {\boldsymbol f_1},\, \mathbf{\Delta}_1 \rangle_{\rm F} \\ =& \langle \mathbf{A}_0,\, \mathbf{\Delta}_0 \rangle_{\rm F} + \langle \mathbf{A}_1,\, \mathbf{\Delta}_1 \rangle_{\rm F}\;,
    \end{aligned}
    \label{eq:prop_example_}
\end{equation}
where $\mathbf{A}_t$ is the architectural matrix for layer $t$, which is AM-independent and only depends on the original exact CNN.

Typically, for an AM-CNN containing $T$ approximate layers, the error expectation of the last approximate layer is
\begin{equation}
    {\rm E}(e_{last}) \approx \sum_{t=1}^{T} \langle \mathbf{A}_t ,\, \mathbf{\Delta}_t \rangle_{\rm F}\;,
    \label{eq:last_E}
\end{equation}

In the case where all approximate layers use the same type of AMs, \refeq{eq:last_E} can be transformed to
\begin{equation}
    \text{AME} = \langle \sum_{t=1}^{T}\mathbf{A}_t ,\, \mathbf{\Delta} \rangle_{\rm F}\;.
    \label{eq:AME}
\end{equation}
As \refeq{eq:AME} has the same form as \refeq{eq:ME_matrix}, we consider it as a new AM error metric, referred to as the architectural mean error (AME). AME contains two aspects of information that are crucial to the AM-CNN accuracy, i.e., the CNN architecture ($\mathbf{A}$) and the AM error characteristics ($\mathbf{\Delta}$). These two elements are mutually independent, thus can be obtained separately.

Note that \refeq{eq:last_E} is feasible to guide layer-wise approximation. However, to reduce the search space size in this work, we prefer to employ the AME to facilitate the selection or design of a single appropriate AM. This is consistent with the scenario of arithmetic circuit reuse in resource-constrained hardware platforms.

\section{Experiments}
\label{sec:exp}
\begin{figure}[t]
    \begin{center}
        \includegraphics [width=1.0\linewidth]{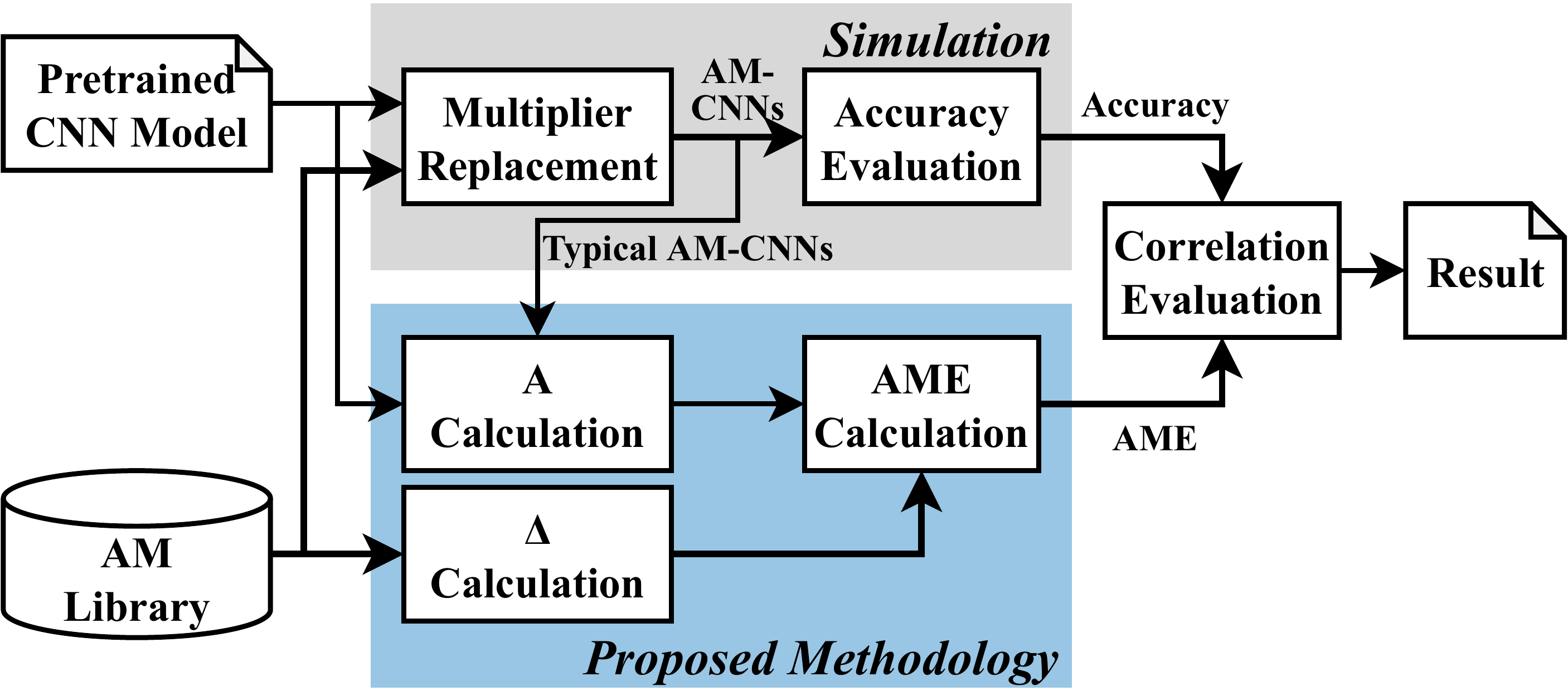} 
        \caption{Overview of the experiments.}
        \label{fig:eva_flow}
    \end{center}
\end{figure}

In this section, the effectiveness of the proposed AME is evaluated by the experiments illustrated in \reffig{fig:eva_flow}. The pretrained CNN models include VGG-16 on CIFAR-10 and ImageNet, ResNet-18 on CIFAR-10 and CIFAR-100, ResNet-34 on CIFAR-10 and CIFAR-100, and ResNet-50 on ImageNet. We focus on the $8\times 8$ unsigned AM because it is one of the most commonly used AM configurations in AM-CNNs~\cite{mrazek2019alwann, mo2023learning}.

The AM library contains four categories of $8 \times 8$ unsigned AMs, with a total of 1,481 AMs.
\begin{itemize}
    \item 500 AMs were taken from EvoApprox8b~\cite{mrazek2017evoapprox8b}.
    \item 352 AMs were generated by removing some full adders from an exact Wallace tree multiplier~\cite{mo2023learning}.
    \item 315 AMs were generated by replacing some full adders in an exact Wallace tree multiplier with approximate full adders~\cite{mo2023learning}.
    \item 314 manually designed AMs found in literatures involving in various approximation techniques, such as dynamic range detection~\cite{hashemi2015drum}, logarithmic approximation~\cite{SOA18, ILM20}, and using approximate compressors for the partial product accumulation~\cite{CP20, CP23}.
\end{itemize}

When constructing an AM-CNN, all the multipliers in the convolutional layers of a pretrained CNN model are replaced with the same type of AMs selected from the AM library. To prevent unexpected accuracy degradations, the other arithmetic operations and the convolutional layers with the kernel size equal to 1 in ResNet remain accurate. To enable the multiplications to be performed by using $8\times 8$ unsigned multipliers, the input activations and weights of the convolutional layers were quantized to 8-bit unsigned integers.

\begin{table*}[t]
    \caption{PCCs between the classification accuracy of AM-CNNs and the AM error metrics.}
    \begin{center}
        \scalebox{1.0}{
        \begin{tabular}{ccccccccccc}
            \hline
            \multirow{2}*{\textbf{CNN}} & \multirow{2}*{\textbf{Dataset}} & \multirow{2}*{\textbf{\tabincell{c}{VarE \\ \cite{ansari2019improving}}}} &
            \multirow{2}*{\textbf{\tabincell{c}{RMSE \\ \cite{ansari2019improving}}}} &
            \multirow{2}*{\textbf{\tabincell{c}{VarRE \\ \cite{kim2021effects}}}} &
            \multicolumn{2}{c}{\textbf{ME \cite{mo2023learning}}} &
            \multirow{2}*{\textbf{\tabincell{c}{MED \\ \cite{mo2023learning}}}} &
            \multirow{2}*{\textbf{\tabincell{c}{ER \\ \cite{mo2023learning}}}} &
            \multicolumn{2}{c}{\textbf{AME}}\\
            \cline{6-7}  \cline{10-11}
            ~ & ~ & ~ & ~ & ~ & \textbf{Neg} & \textbf{Pos} & ~ & ~ & \textbf{Neg} & \textbf{Pos}\\
            \hline
            VGG-16 & CIFAR-10 & -0.177 & -0.168 & -0.353 & 0.224 & -0.279 & -0.268 & -0.634 & \textbf{0.840} & \textbf{-0.770} \\ 
            VGG-16 & ImageNet & -0.016 & -0.143 & -0.093 & 0.118 & 0.124 & -0.129 & -0.383 & \textbf{0.919} & \textbf{-0.863} \\ 
            ResNet-18 & CIFAR-10 & 0.051 & 0.026 & -0.384 & 0.123 & -0.894 & -0.186 & -0.384 & \textbf{0.838} & \textbf{-0.898} \\
            ResNet-18 & CIFAR-100 & 0.121 & 0.123 & -0.260 & 0.302 & \textbf{-0.912} & -0.318 & -0.461 & \textbf{0.908} & -0.910 \\
            ResNet-34 & CIFAR-10 & 0.089 & 0.016 & -0.258 & 0.577 & -0.745 & -0.554 & -0.452 & \textbf{0.853} & \textbf{-0.791} \\
            ResNet-34 & CIFAR-100 & 0.150 & 0.177 & -0.072 & 0.209 & \textbf{-0.887} & -0.315 & -0.510 & \textbf{0.937} & -0.743 \\
            ResNet-50 & ImageNet & -0.036 & -0.022 & -0.065 & 0.136 & -0.015 & -0.107 & -0.438 & \textbf{0.873} & \textbf{-0.837} \\
            \hline
        \end{tabular}}
    \label{tab:AME_cor}
    \end{center}
\end{table*}

\begin{figure}[t]
    \begin{center}
        \includegraphics [width=1.0\linewidth]{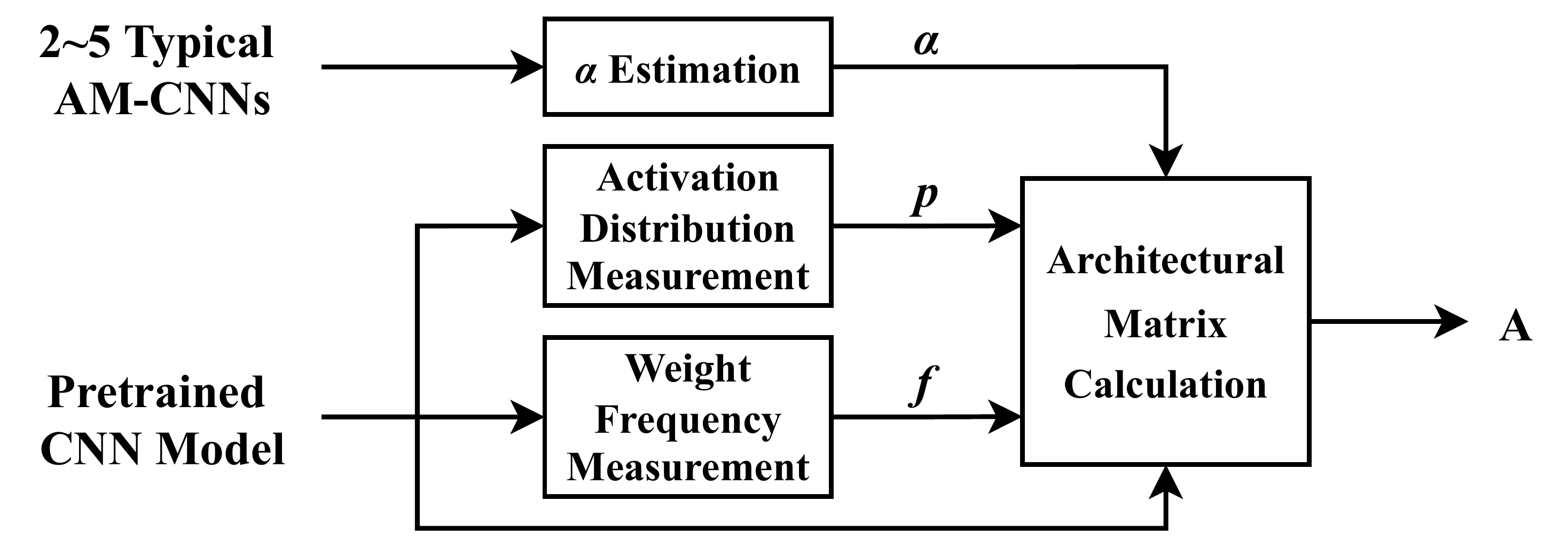} 
        \caption{Calculation process of the architectural matrix $\mathbf{A}$.}
        \label{fig:calculate_A}
    \end{center}
\end{figure}

\reffig{fig:calculate_A} shows the calculation process of the architectural matrix $\mathbf{A}$. Specifically, 2 to 5 typical AM-CNNs corresponding to each pretrained CNN model are selected, and the error propagation factor $\alpha$ is estimated using the method described in \refsec{subsubsec:conv}. ${\boldsymbol p}$ and ${\boldsymbol f}$ were obtained by conducting data statistics on the pretrained CNN model using a dataset comprising images sampled from the training dataset. In this work, we sampled 10,000 images for CIFAR-10 and CIFAR-100, and 1,280 images for ImageNet.

The error matrix $\mathbf{\Delta}$ of an AM was obtained through traversing all possible input combinations and recording the output errors. Specifically, if the LUT of the AM is known, only one subtraction from the LUT of the exact multiplier is required to obtain the $\mathbf{\Delta}$.

The experiments were conducted on Intel Xeon Gold 6226R CPU and NVIDIA GTX 3090 GPU. The AM-CNN simulations were based on the TFApprox~\cite{vaverka2020tfapprox}, which is a TensorFlow-based framework that supports GPU-based simulation.

\subsection{Correlation Evaluation}
\label{subsec: Correlation}

By using \refeq{eq:pcc}, the PCC was calculated to evaluate the correlation between the AME and the accuracy of AM-CNNs. In addition, several AM error metrics that have been proved to be important to the AM-CNN accuracy from recent studies were considered for comparison. The evaluation included only AM-CNNs with less than 30\% accuracy degradation, as those exceeding this threshold are commonly unacceptable.

\reftab{tab:AME_cor} shows the PCCs between the AM-CNN accuracy and the AM error metrics. Overall, the correlation between the accuracy and the AME is quite strong, especially for the negative AME. The positive ME performs slightly better than positive AME in two applications (ResNet-18 on CIFAR-100 and ResNet-34 on CIFAR-100); however, it performs poorly in other applications, especially for those with complex dataset (i.e., ImageNet).
Therefore, AME is shown to be a generic AM error metric that strongly correlates with the AM-CNN accuracy. The universality of AME arise from the information it contains about the CNN architecture.

Furthermore, using AME as the input, we trained regression models to predict the accuracy of AM-CNNs with less than 30\% accuracy degradation. The ME, MED, and ER that perform relatively well in \reftab{tab:AME_cor} were considered for comparison. In the training, the quadratic regression algorithm was utilized, which makes the model simpler than most state-of-the-art learning-based designs~\cite{mrazek2019autoax, ansari2019improving, ullah2022appaxo, mo2023learning}. The evaluation metric of the regression model is the mean absolute percentage error (MAPE), given by
\begin{equation}
    \text{MAPE} = \frac{1}{N_s}\sum_{i=1}^{N_s} |\frac{c_i-\hat{c_i}}{c_i}|\times 100\%\;,
    \label{eq:mape}
\end{equation}
where $N_s$ is the number of samples, $\hat{c_i}$ and $c_i$ are the predicted and actual accuracy of the $i$th AM-CNN, respectively.

\begin{table}[t]
    \caption{MAPEs (\%) of regression models.}
    \begin{center}
        \scalebox{1.0}{
        \begin{tabular}{ccccccccc}
            \hline
            \textbf{CNN} & \textbf{Dataset} & \textbf{ME} & \textbf{MED} & \textbf{ER} & \textbf{AME}\\
            \hline
            VGG-16 & CIFAR-10 & 6.242 & 6.230 & 5.267 & \textbf{1.950} \\ 
            VGG-16 & ImageNet & 11.76 & 11.36 & 9.345 & \textbf{6.331} \\ 
            ResNet-18 & CIFAR-10 & 5.336 & 6.391 & 8.330 & \textbf{3.498} \\
            ResNet-18 & CIFAR-100 & 5.965 & 6.710 & 7.274 & \textbf{3.287}\\
            ResNet-34 & CIFAR-10 & 5.445 & 6.190 & 7.286 & \textbf{4.848} \\
            ResNet-34 & CIFAR-100 & 6.524 & 5.564 & 8.443 & \textbf{3.706} \\
            ResNet-50 & ImageNet & 7.790 & 7.835 & 7.444 & \textbf{3.539} \\ 
            \hline
        \end{tabular}}
    \label{tab:predictor}
    \end{center}
\end{table}

\begin{figure}[t]
        \subfigure[ME\label{fig:resnet50imagenet_ME}]
	{\includegraphics[width=0.24\textwidth]{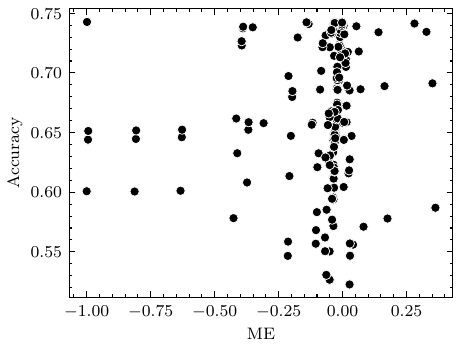}}
        \subfigure[MED\label{fig:resnet50imagenet_MED}]
	{\includegraphics[width=0.24\textwidth]{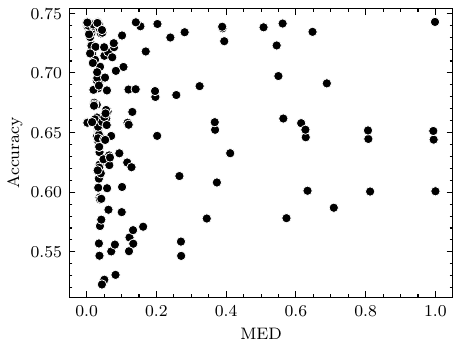}}
         \subfigure[ER\label{fig:resnet50imagenet_ER}]
	{\includegraphics[width=0.24\textwidth]{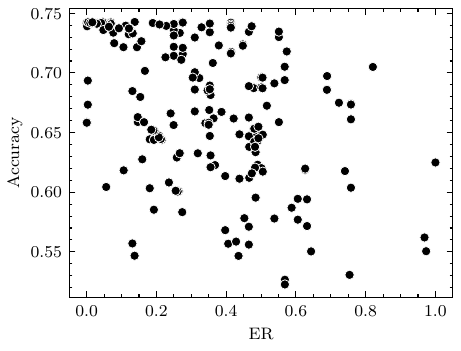}}
        \subfigure[AME\label{fig:resnet50imagenet_AME}]
	{\includegraphics[width=0.24\textwidth]{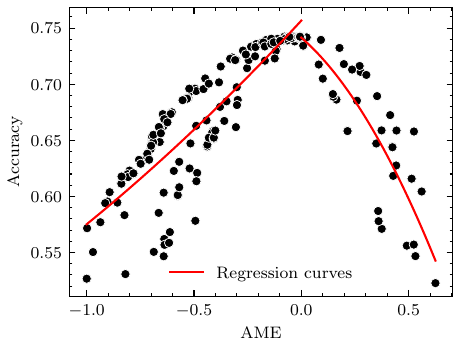}}
	\caption{Scatterplots of AMs, where the abscissa is the normalized error metric of the AM and the ordinate is the classification accuracy of the corresponding AM-based ResNet-50 on ImageNet.}
	\label{fig:predictor}
\end{figure}

\reftab{tab:predictor} shows the MAPEs of the obtained regression models. It can be seen that the AME-based models exhibit high prediction accuracy, outperforming other models in all applications. 
Taking the ResNet-50 on ImageNet as an example, \reffig{fig:predictor} more intuitively demonstrates the performance of AME in the accuracy prediction. Compared to other metrics (ME, MED, and ER), the relationship between AME and accuracy is clearer, allowing it to be effectively modeled with quadratic regression curves\footnote{Due to the differing correlations between positive and negative AME and accuracy, two separate regression curves were employed to handle positive and negative AME respectively, as shown in \reffig{fig:resnet50imagenet_AME}.}.

\subsection{Time Requirements}
We compare the time requirement of the AME-based prediction with that of the TFApprox-based simulation.

Taking the VGG-16 on ImageNet as an example, without considering parallelism, \reftab{tab:time} presents the average time consumed by an AM-CNN on each major step of AME evaluation. It shows that calculating $\mathbf{A}$ is the most time-consuming step, which includes the layer-wise data statistics for the pretrained CNN model and typical AM-CNNs. Specifically, it takes 1,702 seconds to measure the error propagation factors ($\alpha$) of all layers in an AM-CNN, and this measurement was performed on four AM-CNNs to determine credible $\alpha$. In addition, it takes 3628 seconds to obtain the input activation distributions and weight frequencies for all convolutional layers in the pretrained CNN model.

\begin{table}[t]
    \caption{Average time spent per step for evaluating the AME of an AM-based VGG-16 on ImageNet.}
    \begin{center}
        \scalebox{1.0}{
        \begin{tabular}{lr}
            \hline
            \textbf{Step} & \textbf{Time/s}\\
            \hline
            $\mathbf{A}$ Calculation & $1702 \times 4 + 3628 = 10,436$\\ 
            $\mathbf{\Delta}$ Calculation & $1.9\times10^{-5}$\\
            AME Calculation & $1.0\times10^{-5}$\\
            \hline
        \end{tabular}}
    \label{tab:time}
    \end{center}
\end{table}

However, the $\mathbf{A}$ calculation only needs to be performed once for a CNN model. After determining the $\mathbf{A}$, only two steps are required to obtain the AME of an AM, i.e., calculating the $\mathbf{\Delta}$ followed by a Frobenius inner product of $\mathbf{A}$ and $\mathbf{\Delta}$. Therefore, when the architecture of the CNN is given, the required time for the AME-based prediction is on the order of $10^{-5}$s. \reftab{tab:speedup} shows the speedup of AME-based predictions compared to TFApprox-based simulations in various applications. The geometric mean speedup over these applications is about $2\times10^{6}$.

\begin{table}[t]
    \caption{Speedup of AME-based prediction compared to TFApprox-based simulation.}
    \begin{center}
        \scalebox{0.95}{
        \begin{tabular}{ccccc}
            \hline
            \textbf{CNN} &\textbf{Dataset} & \textbf{\tabincell{c}{Simulation \\ Time/s}} & \textbf{\tabincell{c}{AME Evaluation \\ Time/s}} & \textbf{Speedup}\\
            \hline
            VGG-16 & CIFAR-10 & $9$ & \bm{$2.9\times10^{-5}$} & $3.1 \times 10^{5}$ \\ 
            VGG-16 & ImageNet & $1286$ & \bm{$2.9\times10^{-5}$}& $4.4 \times 10^{7}$ \\ 
            ResNet-18 & CIFAR$^{\mathrm{a}}$ & $16$ & \bm{$2.9\times10^{-5}$} & $5.5 \times 10^{5}$ \\
            ResNet-34 & CIFAR$^{\mathrm{a}}$ & $30$ & \bm{$2.9\times10^{-5}$} &  $1.0 \times 10^{6}$\\
            ResNet-50 & ImageNet & $205$ & \bm{$2.9\times10^{-5}$} & $7.1 \times 10^{6}$ \\ 
            \hline
            \multicolumn{3}{l}{$^{\mathrm{a}}$CIFAR-10 or CIFAR-100.}
        \end{tabular}}
    \label{tab:speedup}
    \end{center}
\end{table}

\section{Application of the Proposed AME}
\label{sec:application}

A typical application of the proposed AME is to expedite the AM selection process for CNN applications. As an example, we used AME to select the Pareto-optimal AMs from the EvoApprox8b library~\cite{mrazek2017evoapprox8b} for CNN-based image classification applications.

EvoApprox8b contains 500 $8 \times 8$ unsigned AMs. This experiment aims to pick out the AMs optimized for the area\footnote{Areas of AMs in this experiment were determined through synthesis using Synopsys Design Compiler with 28nm CMOS technology.} of the AM and the accuracy of the corresponding AM-CNN. To avoid time-consuming simulations, we utilized the AME instead of AM-CNN accuracy as an optimization objective. In other words, for AMs with the same area, the one with the lowest absolute value of AME is selected.

As the AM-CNN accuracy predicted based on AME slightly deviates from the actual value, we introduced an iterative search method to prevent missing the real Pareto-optimal AMs, as shown in Algorithm~\ref{alg:search}. Specifically, at each iteration, the Pareto-optimal AM set with respect to area and AME is obtained, these AMs are recorded and removed from the library. This process is repeated until the iteration number reaches the preset threshold. Finally, a pseudo Pareto-optimal AM set is constructed by merging the AM sets obtained in all iterations. Thus, a larger iteration number is likely to result in more AMs. The Pareto-optimal set search step (line 3) in Algorithm~\ref{alg:search} is a multi-objective optimization problem, for which a traversal-based method was utilized in this experiment. Since the searching method does not affect the experimental results, its details are omitted.

\renewcommand{\algorithmicensure}{\text{Outputs:}} 
\renewcommand{\algorithmicrequire}{\text{Inputs:}}  
\begin{algorithm}
	\caption{Iterative Search Algorithm} 
	\label{alg:search}
	\begin{algorithmic}[1]
		\REQUIRE AM library $D$; Total number of iterations $n$
		\ENSURE Pseudo Pareto-optimal AM set $D^*$
	       
		\STATE $D^* = \emptyset$, $B = D$, $i = 0$
		
		\WHILE{$i < n$} 
		\STATE $P = SearchParetoOptimalSet(B)$ 
            \STATE $D^* = D^* \cup P$ 
		\STATE $B = B \setminus P$
		\STATE $i=i+1$
		\ENDWHILE
	\end{algorithmic}
\end{algorithm}

\begin{figure*}[t]
	\centering
        \subfigure[VGG-16 on CIFAR-10 (after 6 iterations)\label{fig:vgg16cifar10_app}]
	{\includegraphics[width=0.32\linewidth]{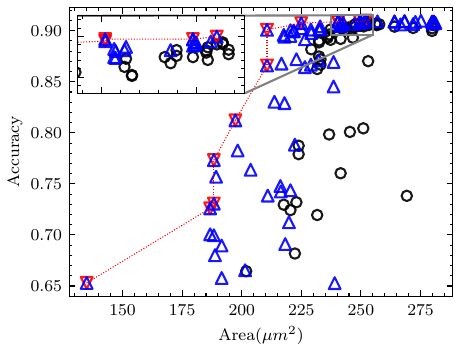}}
        \subfigure[VGG-16 on CIFAR-10 (after 3 iterations)\label{fig:vgg16cifar10_app_3}]
	{\includegraphics[width=0.32\linewidth]{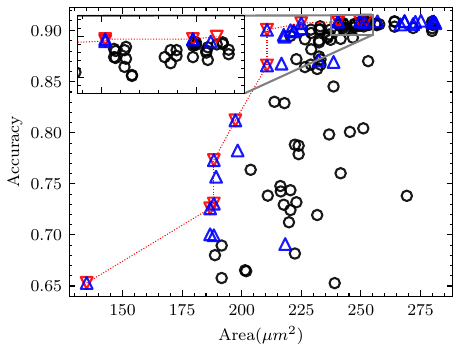}}
        \subfigure[VGG-16 on ImageNet\label{fig:vgg6imagenet_app}]
	{\includegraphics[width=0.32\linewidth]{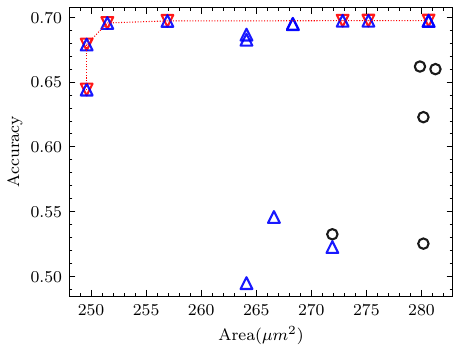}}
	\subfigure[ResNet-18 on CIFAR-10\label{fig:resnet18cifar10_app}]
	{\includegraphics[width=0.32\linewidth]{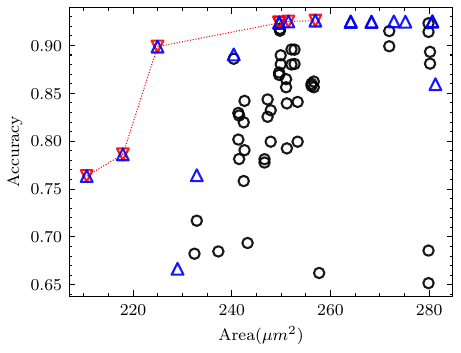}}
	\subfigure[ResNet-18 on CIFAR-100\label{fig:resnet18cifar100_app}]
	{\includegraphics[width=0.32\linewidth]{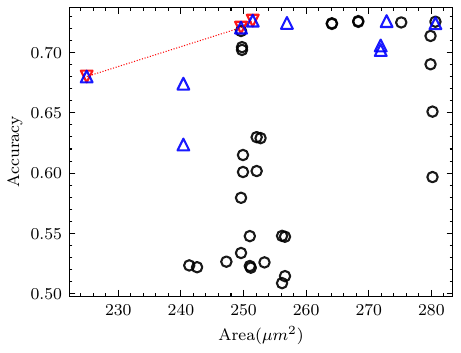}}
        \subfigure[ResNet-34 on CIFAR-10\label{fig:resnet34cifar10_app}]
	{\includegraphics[width=0.32\linewidth]{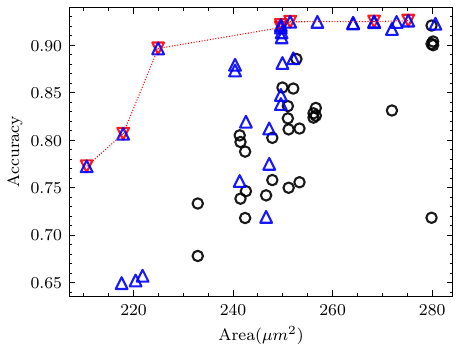}}
	\subfigure[ResNet-34 on CIFAR-100\label{fig:resnet34cifar100_app}]
	{\includegraphics[width=0.32\linewidth]{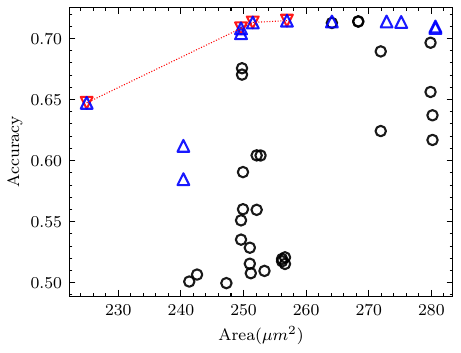}}
        \subfigure[ResNet-50 on ImageNet\label{fig:resnet50imagenet_app}]
	{\includegraphics[width=0.32\linewidth]{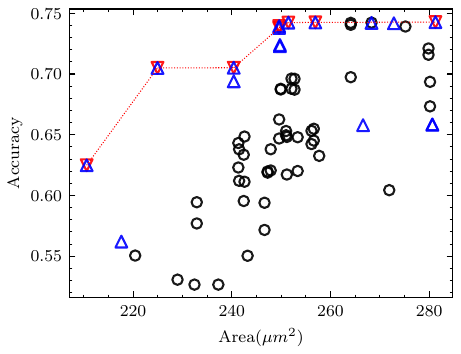}}
        \begin{minipage}[t]{0.32\textwidth}
		\centering
		\includegraphics[width=0.9\textwidth]{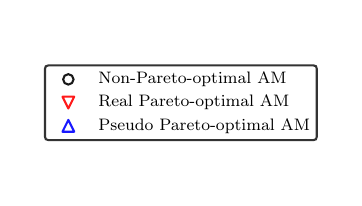}
	\end{minipage}
	\caption{AM selection for CNN-based image classification applications. AM-CNNs with accuracy degradation exceeding 30\% are omitted from these figures for clarity.}
	\label{fig:application}
\end{figure*}

For each CNN application in the experiment, we measured the minimum number of iterations required to ensure that the pseudo Pareto-optimal AM set encompasses all  real Pareto-optimal AMs with accuracy degradation not exceeding 30\%, as shown in \reftab{tab:application}. As observed, for applications other than VGG-16 on CIFAR-10, 2 iterations suffice to ensure no real Pareto-optimal AM is missed. In fact, VGG-16 on CIFAR-10 has the best error tolerance among these applications, many AMs do not cause a significant accuracy degradation, as shown in \reffig{fig:vgg16cifar10_app}. In this case, the deviation of AME-based prediction has a relatively significant impact. However, as shown in \reffig{fig:vgg16cifar10_app_3}, after 3 iterations, we have covered all real Pareto-optimal AMs except the one with the largest area and the highest AM-CNN accuracy. This is adequate for AM-CNN designs where minimizing area is the primary optimization objective. Moreover, since the time required to compute an AME is on the order of $10^{-5}$s, the supplementary time overhead incurred by multiple iterations is negligible.

\begin{table}[t]
    \caption{The minimum number of iterations and the proportion of real Pareto-optimal AMs in the results.}
    \begin{center}
        \scalebox{1.0}{
        \begin{tabular}{cccc}
            \hline
            \textbf{CNN} & \textbf{Dataset} & \textbf{\tabincell{c}{Iteration \\ Number}} & \textbf{Proportion}\\
            \hline
            VGG-16 & CIFAR-10 & 6 & 12/80 \\ 
            VGG-16 & ImageNet & 2 & 7/15 \\ 
            ResNet-18 & CIFAR-10 & 2 & 7/19 \\
            ResNet-18 & CIFAR-100 & 1 & 3/10 \\
            ResNet-34 & CIFAR-10 & 2 & 8/31 \\
            ResNet-34 & CIFAR-100 & 2 & 4/12 \\
            ResNet-50 & ImageNet & 2 & 8/17 \\ 
            \hline
        \end{tabular}}
    \label{tab:application}
    \end{center}
\end{table}

The proportion of the real Pareto-optimal AMs in the final pseudo Pareto-optimal AM set was also recorded, as shown in \reftab{tab:application}. \reffig{fig:application} provides a more intuitive representation of the experimental results. It shows that the real Pareto-optimal AMs are encompassed within the pseudo Pareto-optimal AMs selected based on AME. To distinguish the real Pareto-optimal AMs, subsequent simulations can be performed within a much less number of AMs. This is much faster than searching directly in the library containing hundreds of AMs.

Take the VGG-16 on ImageNet as an example. According to \reftab{tab:time} and \reftab{tab:speedup}, the total time spent evaluating the AMEs corresponding to 500 AMs is about $10,436+2.9\times10^{-5}\times500 \approx 10,436$ seconds, which is comparable to the time spent simulating the accuracy of 9 AM-CNNs ($1,286 \times 9 = 11,574$ seconds). Hence, the initial AM selection based on AME, followed by simulation and filtering, results in a total time requirement approximately equivalent to performing $9+15=24$ simulations of AM-CNNs. This approach is roughly 21$\times$ faster than the traditional method of simulating 500 AM-CNNs. It is anticipated that as the number of AMs in the library grows, the speedup ratio will likewise increase.

This experiment illustrates that, given a search space, AME-based selection can quickly prune it, thereby speeding up the search for the Pareto-optimal AMs for CNN applications.

Furthermore, AME has the potential to facilitate the automated design of CNN-oriented AM. In contrast to selecting pre-existing generic AM from a library, automated design aims to generate the optimal hardware for AMs under specific constraints.~\cite{hrbacek2016automatic, ullah2022appaxo, li2022adaptable}. The process of automated design typically includes circuit coding, design space construction, and design space search. For example, the AMs of EvoApprox8b are automated designed through a Cartesian genetic programming (CGP)-based methodology~\cite{hrbacek2016automatic}. Specifically, a candidate AM is represented as a fixed-sized 2D Cartesian grid of nodes interconnected by a feed-forward network, where each node represents a 2-input Boolean function (e.g., AND, OR). Consequently, a candidate solution can be represented by a netlist that records the node connections. After determining candidate netlists (i.e., the design space), a multi-objective evolutionary algorithm was executed to search for the Pareto-optimal AM set with respect to error metrics and hardware costs.

It is noteworthy that the error metric used as the optimization objective during the design space search process determines the error characteristics of the AM obtained. When utilizing an application-independent AM error metric, such as the mean relative error distance (MRED)~\cite{hrbacek2016automatic}, the resulting AMs are also application-independent. However, when utilizing an AM error metric that incorporates the data distribution of a specific CNN application~\cite{li2022adaptable}, the obtained AMs tend to maintain a high accuracy for the corresponding AM-CNNs. As the proposed AME demonstrates a strong correlation with the AM-CNN accuracy and can be efficiently evaluated, it holds significant potential to serve as an optimization objective in the automated design of AMs for CNN applications.

\section{Discussion}
\label{sec:discussion}
In this section, the rationale behind the effectiveness of the proposed AME is further discussed.

\begin{figure}[t]
        \subfigure[VGG-16 on CIFAR-10\label{fig:cmp_vgg16}]
        {\includegraphics[width=0.49\textwidth]{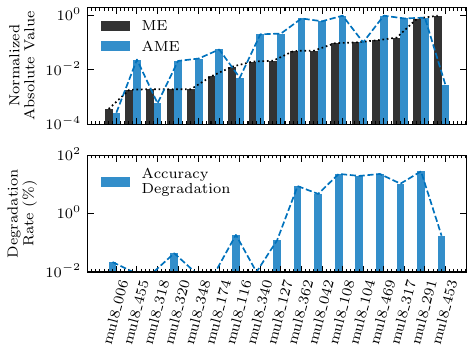}}\\
         \subfigure[ResNet-50 on ImageNet\label{fig:cmp_resnet18}]
        {\includegraphics[width=0.49\textwidth]{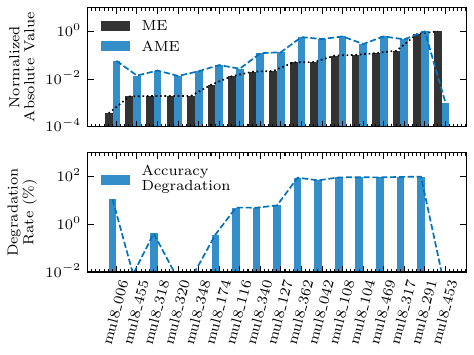}}
	\caption{Histograms of the normalized absolute values of AM error metrics and the accuracy degradations of the corresponding AM-CNNs.}
	\label{fig:cmp}
\end{figure}

As per the experimental results shown in \refsec{subsec: Correlation}, among the error metrics highlighted in existing researches (i.e., VarE~\cite{ansari2019improving}, RMSE~\cite{ansari2019improving}, VarRE~\cite{kim2021effects}, and ME~\cite{mo2023learning}), ME shows the strongest correlation with the accuracy of AM-CNNs. Thus, ME is visually compared with AME, as shown in \reffig{fig:cmp}. \reffig{fig:cmp} presents the values of ME, AME, and accuracy degradation in two CNN applications for AMs belonging to the union of the real Pareto-optimal AM sets obtained in \refsec{sec:application}. The AMs are listed in an ascending order of their absolute values of ME.

As observed in \reffig{fig:cmp}, AME has the following two superiorities over ME.
\begin{itemize}
    \item \textbf{AME exhibits a trend closely aligned with the accuracy degradation of AM-CNNs.} For example, the mul8\_453 has the maximum absolute value of ME among these AMs; however, it only slightly degrades the accuracy of the two CNN applications. The AMEs corresponding to mul8\_453 are significantly low. Hence, AME is more suitable to capture the impact of this AM on the accuracy of AM-CNNs.
    \item \textbf{AME varies depending on the application.} For instance, the mul8\_006 has the minimum absolute value of ME among these AMs, and it slightly degrades the accuracy of the VGG-16 on CIFAR-10. However, when applied to ResNet-50 on ImageNet, mul8\_006 leads to an accuracy degradation of more than 10\%. The AME of mul8\_006 is low for VGG-16 on CIFAR-10, whereas it is much higher for ResNet-50 on ImageNet. In this case, AME can predict the accuracy degradation but ME cannot.
\end{itemize}

\begin{figure}[t]
         \subfigure[Weights of layer 1\label{fig:weight_dist0}]
        {\includegraphics[width=0.49\linewidth]{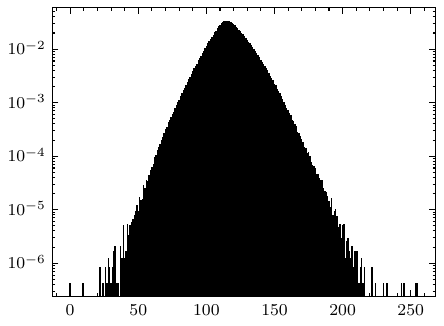}}
        \subfigure[Input activations of layer 1\label{fig:activation_dist0}]
        {\includegraphics[width=0.49\linewidth]{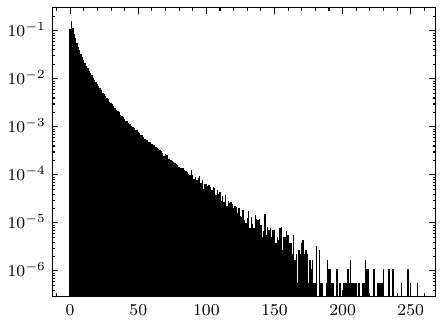}}\\
        \subfigure[Weights of layer 2\label{fig:weight_dist1}]
	{\includegraphics[width=0.49\linewidth]{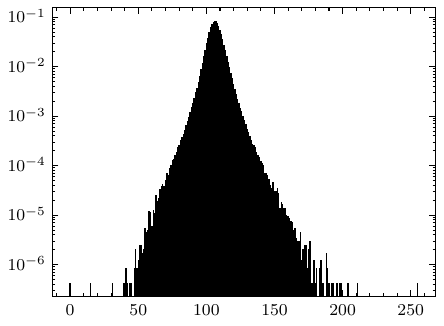}}
        \subfigure[Input activations of layer 2\label{fig:activation_dist1}]
	{\includegraphics[width=0.49\linewidth]{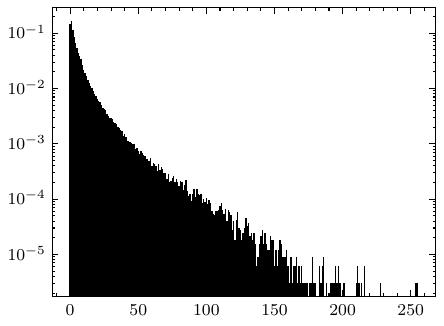}}
	\caption{Histograms of the quantized weights and input activations of the last two convolutional layers in VGG-16 on CIFAR-10.}
	\label{fig:dist}
\end{figure}

Compared to the calculation formula for ME (given by \refeq{eq:ME_matrix}), the formula for AME (given by \refeq{eq:AME}) replaces the data distribution matrix $\mathbf{D}$ with the summation of architectural matrices $\mathbf{A}$. The matrix $\mathbf{D}$ is application-independent, and generally represents a joint uniform distribution (i.e., all elements in $\mathbf{D}$ are equal). In fact, most researches on AM error metrics are conducted under the assumption that the inputs of the AM follow a joint uniform distribution~\cite{ansari2019improving, mo2023learning, kim2021effects}. However, the inputs to the multipliers in the convolutional layer are not uniformly distributed. For example, \reffig{fig:dist} shows the data distributions of the last two convolutional layers in VGG-16 on CIFAR-10, where the weights and the input activations follow significantly disparate distributions. Hence, the ME calculated under the assumption of uniformly distributed inputs may not align with the actual application scenario. Some state-of-the-art researches have acknowledged this issue and measured data distributions from specific CNN applications to guide the design of CNN-oriented AMs~\cite{guo2020reconfigurable, li2022adaptable}. 

Comparing \reffig{fig:weight_dist0} with \reffig{fig:weight_dist1}, and \reffig{fig:activation_dist0} with \reffig{fig:activation_dist1}, it can be seen that even the data distributions of two adjacent layers can be different. According to \refeq{eq:prop_example_}, the architectural matrix of a layer can be viewed as the weighted data distribution of the layer, i.e., ${\boldsymbol p}^\mathrm{T} {\boldsymbol f}$ multiplied by a constant. Therefore, the summation of architectural matrices can be considered as a fusion of the data distributions of all approximate layers. This fusion is based on the proposed linear model for fitting the error propagation process in AM-CNNs. Consequently, AME contains more information about applications than traditional AM error metrics, thereby performing better in accuracy predictions for AM-CNNs.

\section{Conclusion}
\label{sec:conclusion}
In this work, the error generation and propagation process within AM-CNNs is linearly modeled. Based on this model, a novel AM error metric, the architectural mean error (AME), is proposed to evaluate the accuracy degradation of an AM-CNN. The proposed AME involve the information of CNN architectures and AM error characteristics, thus exhibiting a strong correlation with the AM-CNN accuracy. Experimental results on various CNN models and datasets show that the AME can effectively predict AM-CNN accuracy with rapidity and high precision. Therefore, AME can be employed to facilitate the AM selection and design for CNN applications.

\bibliographystyle{IEEEtran}
\bibliography{ErrorMetric}
\end{document}